# Improved Photoelectrochemical Water Splitting of CaNbO$_2$N Photoanodes by Co-Pi Photodeposition and Surface Passivation

Fatima Haydous,[†] Wenping Si,[†§] Vitaliy A. Guzenko,[‡] Friedrich Waag,[⊥] Ekaterina Pomjakushina,[†] Mario El Kazzi,[∥] Laurent Sévery,[•] Alexander Wokaun,[∥] Daniele Pergolesi*,[†,∥] and Thomas Lippert*,[†,#,∇]

[†] Division for Research with Neutrons and Muons, Paul Scherrer Institut, 5232 Villigen-PSI, Switzerland

[‡] Photon Science Division, Paul Scherrer Institut, 5232 Villigen-PSI, Switzerland

[⊥] Center for Nanointegration Duisburg-Essen, Technical Chemistry I, University of Duisburg-Essen

[∥] Energy and Environment Research Division, Paul Scherrer Institut, 5232 Villigen PSI, Switzerland

[•] Department of Chemistry, University of Zurich, 8057 Zurich, Switzerland

[#] Laboratory of Inorganic Chemistry, Department of Chemistry and Applied Biosciences, ETH Zurich, 8093 Zurich, Switzerland

[∇] Molecular Photoconversion Devices Division, International Institute for Carbon-Neutral Energy Research (I2CNER) Kyushu University 744 Motooka, 819-0395 Fukuoka, Japan

**ABSTRACT:** Photoelectrochemical solar water splitting is a promising approach to convert solar energy into sustainable hydrogen fuel using semiconductor electrodes. Due to their visible light absorption properties, oxynitrides have shown to be attractive photocatalysts for this application. In this study, the influence of the preparation method of CaNbO$_2$N particles on their morphological and optical properties, and thereby their photoelectrochemical performance, is investigated. The best performing CaNbO$_2$N photoanode is produced by ammonolysis of Nb enriched calcium niobium oxide. The enhanced photoactivity arises from an enlarged surface area and superior visible light absorption properties. The photoactivity of this photoanode was further enhanced by photodeposition of Co-Pi co-catalyst and by atomic layer deposition of an Al$_2$O$_3$ overlayer. A photocurrent density of 70 µA.cm$^{-2}$ at 1.23 V vs RHE was achieved. The observed enhancement of the photoelectrochemical performance after Co-Pi/Al$_2$O$_3$ deposition is the combined effect of the improved kinetics of oxygen evolution due to the Co-Pi co-catalyst and the reduced surface recombination of the photogenerated carriers at the Al$_2$O$_3$ surface layer.

## 1. INTRODUCTION

One of the biggest challenges our world is facing today is the increase in the energy demand associated with the shortage of the fossil fuel supplies. This, in addition to the global warming caused by the carbon dioxide produced from fossil fuels, points at the necessity of finding alternative renewable energy resources. As the sun is capable of supplying Earth with an amount of energy per hour which surpasses the annual global energy need, it stands out to be the most appealing choice among all potential renewable energy sources.[1] The intermittent nature of the solar energy has driven many research efforts towards the conversion of the solar energy into storable and transportable forms.[2] The photoelectrochemical (PEC) water splitting into hydrogen and oxygen using semiconductor electrodes is one of the most promising strategies to efficiently harvest and store the solar energy. Within such an approach, photons in the visible light energy range are absorbed by a semiconducting material and used for the creation of electron-hole pairs. The photogenerated electrons and holes migrate to the surface of the semiconductor where they are used to drive the water reduction and oxidation reaction respectively.

Unfortunately, to date there is not a single photoactive material available that can fulfill all the requirements for a commercially feasible solar hydrogen production process.

The main bottleneck is that most of the materials that are chemically stable in operating conditions are wide band gap metal oxides, with light absorption limited to the ultraviolet spectral region.[3] Many efforts have been devoted to extend the light absorption to the visible light energy range by tuning the metal oxide band gap by doping or sensitizing with dye molecules.[2] On the other hand, for oxides with narrow band gaps, as $Fe_2O_3$ and $WO_3$, the photoactivity is limited by their chemical instability and inappropriate band alignment with water redox potentials.[3]

The narrowed band gap of oxynitrides relative to oxides mainly arises from the negative shift of the valence band maximum due to the hybridization of N2p and O2p orbitals.[4-5] Also the energy position of the conduction band minimum can be affected by the N substitution giving a substantial contribution to the narrowing of the band gap.[6] This characteristic makes oxynitrides very promising photocatalysts for water splitting.

The photoactivity of several oxynitrides was investigated. Currently, $LaTiO_2N$,[7-9] $BaTaO_2N$[10-12] and $TaON$[2, 13-15] are considered to be the best performing materials. Also Nb-based perovskite oxynitrides are expected to achieve good performance mainly due to their band edge structure,[16-17] which is particularly favorable for water splitting. Theoretical calculations[18] suggest for example $CaNbO_2N$ (CNON) as a very promising photocatalyst. It has been demonstrated that this material is in fact capable of both oxidation and reduction of water in the presence of sacrificial agents.[16]

In spite of the good expectations, only a few studies have been conducted so far on this family of oxynitrides which still show inefficient performances.[16-17, 19-20] The main drawback associated with Nb-containing oxynitrides is the ease of reduction of Nb, as compared to Ta for instance, during ammonolysis which results in more defects affecting the optical and conducting properties.[19] In general, the performance of a photocatalyst is limited by different factors including: slow kinetics of hole transfer at the semiconductor's surface to the electrolyte, charge-carrier transport losses, low electron–hole separation rates, surface recombination, and corrosion.[21] The problem of recombination losses has been addressed by facilitating the charge separation and transport and the oxygen evolution reaction (OER) kinetics by loading the photocatalyst with well-known water oxidation catalysts, such as $IrO_2$,[14, 22-23] $CoO_x$,[7, 9, 23-25] and $NiO_x$[23, 26]. $IrO_2$ is one of the most active oxygen evolution co-catalysts, but for large-scale applications co-catalysts consisting of earth-abundant and low-cost elements such as cobalt phosphate (Co-Pi) becomes more appealing.[27] For instance, it was shown that combining $La(Ta,Nb)O_2N$ with Co-Pi co-catalyst improves the charge separation and collection of holes produced at the oxynitride surface, which consequently enhanced its photocatalytic activity.[28]

The recombination losses can also be lowered by reducing the density of defects which improves the crystal quality. Concerning Nb-based compounds, it was recently reported that an excess of Nb in the preparation of Ba and Sr niobium oxides was beneficial for the crystallinity and thereby the photoactivity of their corresponding oxynitrides. This was explained considering that the Nb-enriched precursor oxides are isostructural to the corresponding oxynitrides, thus no structural changes occur during the ammonolysis process.[17]

This study provides new insights into the properties and potentials of CNON photoanodes toward solar water splitting. The preparation procedure of the oxynitride is shown to play a key role in determining its PEC performance. We also show here that the photo-assisted deposition of a Co-Pi co-catalyst and the atomic layer deposition (ALD) of a thin passivation layer of $Al_2O_3$ significantly enhance the photoactivity.

## 2. EXPERIMENTAL SECTION

**Synthesis of CaNbO$_2$N Powders.** $CaNbO_2N$ was prepared by both solid state (SS) and polymerized complex (PC) methods.

For the SS reaction, $CaCO_3$ (Alfa Aesar, 99.0% min.) and $Nb_2O_5$ (Merck, 99+%) in stoichiometric amounts were annealed at 1000°C for 12hrs in the presence of a NaCl flux to prepare the $Ca_2Nb_2O_7$ oxide precursor. Then, by thermal ammonolysis of the oxide precursor at 800 °C for 24hrs with mixing in between, $CaNbO_2N$-SS was prepared.

For the PC method, the $Ca_2Nb_2O_7$ oxide precursor was prepared by dissolving stoichiometric amounts of $CaCO_3$ and $NbCl_5$ (ChemPUR, 99+%) in methanol. Then, citric acid (CA) and ethylene glycol (EG) were added with a molar ratio of 1:1:6:2 for Ca:Nb:CA:EG. The solution was left to polymerize at 200°C overnight. Then, the obtained yellow resin was heated at 400°C for 2hrs followed by annealing at 650°C and 800°C for 2hrs each with grinding in between. $CaNbO_2N$-PC was prepared by ammonolysis of the prepared oxide at 800°C for 24hrs.

The PC method was also applied for the preparation of Nb-enriched samples $CaNbO_2N(Nb)$-PC by adding excess Nb up to 1:2 ratio of $CaCO_3$:$NbCl_5$ in the initial oxide mixture.

An ammonia flow of 250mL/min was used for the preparation of all the oxynitrides. The ammonolysis temperature and time were optimized in order to have a single phase of the oxynitride; since at lower temperatures the oxide coexist with the oxynitride and at higher temperatures a $NbO_xN_y$ phase is formed.

**Preparation of Photoanodes.** The photoanodes were prepared via electrophoretic deposition (EPD) where 40 mg of $CaNbO_2N$ powder were mixed with 10 mg of iodine in 50 ml of acetone followed by 1 hour sonication to obtain dispersed $CaNbO_2N$ powders. Electrophoretic deposition (EPD) was conducted between two parallel fluorine-doped tin oxide (FTO) substrates (1×2 cm) placed in the $CaNbO_2N$ dispersion with a distance of 7 mm under a bias of 20 V for 3 min. $CaNbO_2N$ powders were deposited on the negative electrode. A post-necking treatment was done after the deposition of $CaNbO_2N$ on the FTO substrates by dropping 30 μL of 10 mM $TaCl_5$ methanol solution on the $CaNbO_2N$ photoanodes followed by drying in air. After repeating this cycle three times, the photoanodes were annealed at 300 °C for 30 min in air, followed by a heating cycle under ammonia flow for 1 hour at 450 °C.

The Co-Pi co-catalyst was loaded on the $CaNbO_2N$ photoanodes by photodeposition. A solution of 0.5 mM $CoCl_2$ was prepared and 0.1 M potassium phosphate was added to adjust the pH to 7. Afterwards, the sample was dipped in this solution and the Co-Pi co-catalyst was deposited on the $CaNbO_2N$ particles under UV-light illumination for 30 min.

$Al_2O_3$ and $TiO_2$ overlayers were deposited on the Co-Pi loaded $CaNbO_2N$ photoanodes by atomic layer deposition (ALD). For the $Al_2O_3$ layer, successive pulses of trimethylaluminum (TMA) and water vapor were performed in a closed chamber at a temperature of 300 °C with nitrogen as a carrier gas. 20 cycles (TMA/$N_2$ purging/water/$N_2$ purging) were applied resulting in a thickness of around 2 nm for the $Al_2O_3$ layer. For $TiO_2$ layers (about 2 nm thick as well, as measured by ellipsometry on a silicon wafer piece), the deposition was carried out using a Picosun R-200 tool at 120 °C with 32 sequential pulses of tetrakis(dimethylamino)titanium (TDMAT, 99.999% trace metal basis, Sigma-Aldrich) preheated to 85 °C and water vapor.

**Photoelectrochemical (PEC) Measurements.** PEC measurements were performed using a three electrode configuration in 0.5 M NaOH (pH=13.0) aqueous solution with the $CaNbO_2N$ photoelectrode being used as the working electrode. A coiled Pt wire and Ag/AgCl were used as the counter and the reference electrodes, respectively. The electrolyte was purged with Ar for 1 hour prior to PEC measurements. The photoanodes were irradiated from the front side in all the experiments with a 150 W Xe arc lamp (Newport 66477) equipped with AM 1.5 G filter and with an output intensity of 100 mW.cm$^{-2}$ calibrated by a photodetector (Gentec-EO).

**Characterization.** X-ray diffraction (XRD) measurements, conducted by a Bruker–Siemens D500 X-ray Diffractometer, were used to characterize the calcium niobium oxide powders prepared by the solid-state and polymerized-complex routes and their corresponding oxynitrides. The diffuse reflectance of the powders was measured with a Cary 500 Scan UV-Vis-NIR spectrophotometer using an integrating sphere to determine their band gaps. Scanning electron microscopy (SEM) images were obtained with a Zeiss Supra VP55 Scanning Electron Microscope. The surface area was calculated according to the Brunauer–Emmett–Teller (BET) theory by conducting the N2 adsorption–desorption analysis on a Quantachrome Nova 2200 at 77 K. A VG ESCALAB 220iXL spectrometer (Thermo Fischer Scientific) equipped with an Al Kα monochromatic source and a magnetic lens system was used for the X-ray photoelectron spectroscopy (XPS) measurements. Thermogravimetric (TG) measurements were acquired using NETZSCH STA 449C analyzer equipped with PFEIFFER VACUUM ThermoStar mass spectrometer, where aliquots of 20-60 mg of the oxynitride powders were heated in alumina crucibles to 1400 °C with a heating rate of 10 °C /min in 36.8 mL.min$^{-1}$ synthetic air.

## 3. RESULTS AND DISCUSSION

**Characterization of $CaNbO_2N$ and its oxide precursors.** For an efficient PEC water splitting, the optimization of the morphology of the powders is of great importance as it affects the properties of the photocatalyst in terms of active surface area, light absorption, recombination, and charge transfer of the photogenerated carriers.[29] In general, the preparation method affects the morphologies of the resulting oxides and their corresponding oxynitrides. For this purpose, two synthesis routes were used for the preparation of the CNON oxynitrides: the polymerized complex (PC) and solid state (SS) methods.

Figure 1a shows the XRD patterns for the different oxides. The oxide prepared by the SS and PC route starting from stoichiometric ratios of Ca and Nb precursors showed the monoclinic $Ca_2Nb_2O_7$ single phase structure. The PC method was also applied for the preparation of Nb-enriched samples used to probe the effect of the Nb content on the PEC performance. When excess Nb was added in the initial mixture of $CaCO_3$ and $NbCl_5$, the $CaNb_2O_6$ orthorhombic columbite structure was obtained.

CNON was then prepared by thermal ammonolysis at 800°C for 24 hours starting from the $Ca_2Nb_2O_7$ and $CaNb_2O_6$ oxide precursors. Figure 1b shows the XRD patterns of the CNON photoanodes produced from $Ca_2Nb_2O_7$–SS, $Ca_2Nb_2O_7$–PC, $CaNb_2O_6$–PC denoted as CNON-SS, CNON-PC and CNON(Nb)-PC respectively. For all oxynitrides, the orthorhombic perovskite structure was observed, in good agreement with the ICSD reference data, with no evidence of secondary phases. The structures and lattice parameters of the oxides and oxynitrides are reported in Table S1.

We note that the addition of niobium did not result in an oxide isostructural to the oxynitride, as one might have expected by comparison with similar materials. However, a clear difference in the morphologies of the Nb-enriched oxide and oxynitride relative to the other samples is visible in the SEM images.

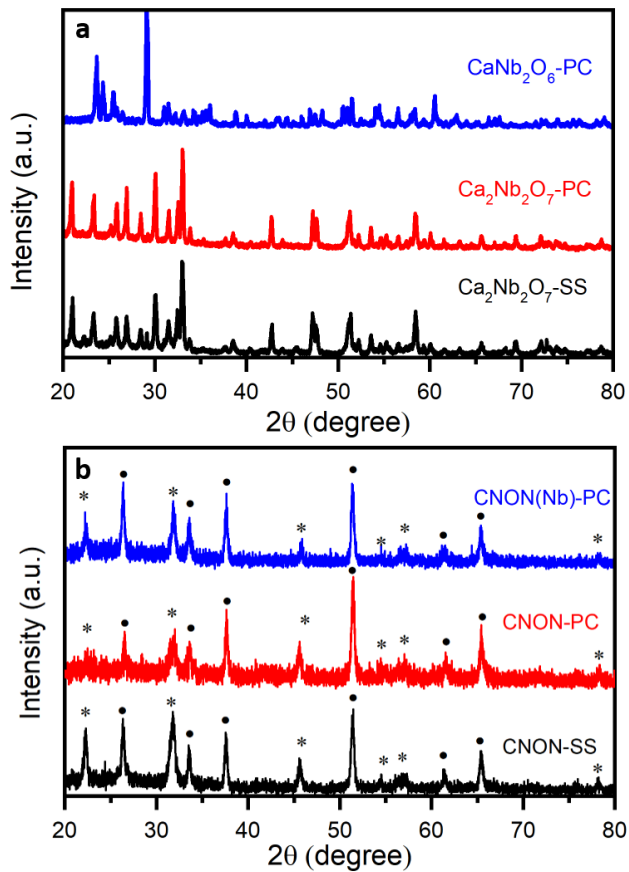

Figure 1. XRD patterns obtained from (a) the calcium niobium oxides and (b) their respective oxynitride photoanodes. The peaks corresponding to the FTO substrate and the perovskite CNON oxynitride are marked with (●) and (∗), respectively.

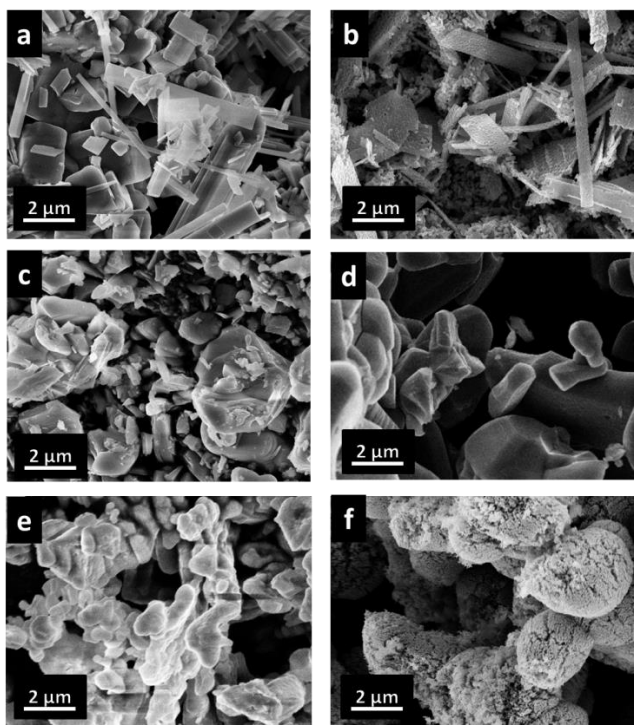

Figure 2. SEM images for (a) $Ca_2Nb_2O_7$-SS (b) CNON-SS (c) $Ca_2Nb_2O_7$-PC (d) CNON-PC (e) $CaNb_2O_6$-PC (f) CNON(Nb)-PC powders.

As shown in Figure 2a, Ca$_2$Nb$_2$O$_7$-SS is obtained in the form of platelets of few μm in size. After ammonolysis, CNON-SS revealed the same brick-like morphology as its original oxide but with a porous structure (Figure 2b). The porosity of the oxynitride is attributed to the exchange of three O$^{2-}$ anions with two N$^{3-}$ anions during the ammonolysis process. A similar morphology has been reported in many studies for LaTiO$_2$N prepared by the solid state route.[30-31] Figure 2c shows the irregularly shaped Ca$_2$Nb$_2$O$_7$-PC particles with a broad size distribution ranging between 0.1 μm and 4 μm. The nitration of this oxide resulted in large irregular aggregated CNON-PC particles (Figure 2d). A different morphology was observed for CaNb$_2$O$_6$-PC, the Nb-enriched oxide obtained by PC method. As shown in Figure 2e, highly aggregated foamy particles with sizes ranging from 0.5 μm to 1.5 μm were obtained for this oxide. After ammonolysis, agglomerates of dense particles are observed for CNON(Nb)-PC with very high porosity (Figure 2f). These agglomerates consist of particles with a small grain size ranging between 50 and 120 nm as shown in Figure S1a.

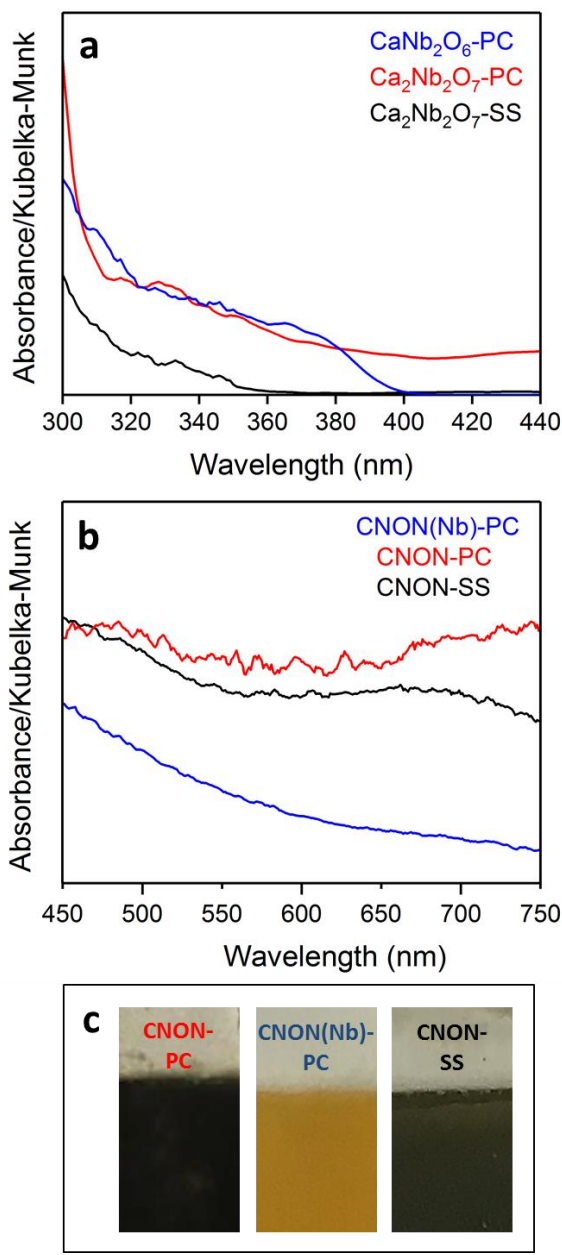

Figure 3. The absorption spectra of (a) the oxides prepared from the solid state and polymerized complex routes (with and without excess of Nb) and (b) their corresponding oxynitrides. The photograph in (c) shows the colors of the three photoanodes.

The absorbance spectra of the three oxynitrides and their corresponding oxides are presented in Figure 3a and b. It can be seen clearly in Figure 3a that $CaNb_2O_6$ has an absorption edge at around 400 nm, more red-shifted compared to $Ca_2Nb_2O_7$ synthesized by the two other routes. In addition, $Ca_2Nb_2O_7$–PC shows a higher absorption background above the light absorption edge compared to the other two oxides which might be attributed to an increased number of defects in this oxide. Crystallographic defects create energy states within the energy gap of the materials thus resulting in absorption at energies smaller than the band gap. Other defects that might be present in $Ca_2Nb_2O_7$–PC are reduced Nb and/or oxygen vacancies which have been shown to increase the background absorption.[17]

After ammonolysis, a red shift in the absorbance is observed for CNON-SS and CNON(Nb)-PC compared to their precursor oxides. The band gaps of these CNON samples were determined from the direct transition in the Tauc plots (Figure S2) to be 1.9 and 2.1 eV for CNON-SS and CNON(Nb)-PC, respectively. For CNON-PC, the Tauc plot in Figure S2b shows a pronounced shoulder at around 2.5 eV which can be attributed to defects inherited from the oxide precursor and/or to the reduced Nb species that contribute to the total absorption. Therefore, the band gap of CNON-PC could not be precisely determined. This effect is also responsible of the high background absorbance shown in Figure 3b which makes impossible to determine the absorption onset of CNON-PC. In general for niobates, the easy reduction of $Nb^{5+}$ during ammonolysis compared to other metal ions (as $Ta^{5+}$)[32] leads to the formation of reduced Nb ($Nb^{3+}$ and $Nb^{4+}$) species which causes an increase in the absorbance background.[17] This also explains the observed high absorbance at wavelengths longer than the absorption edge of CNON-SS. Concerning the Nb-enriched oxynitride, the observed lower background absorbance might be due to the different reactivity of the starting oxide ($CaNb_2O_6$) with $NH_3$ during ammonolysis compared to $Ca_2Nb_2O_7$. Indeed, this is confirmed by the dark brown color of CNON-SS and CNON-PC deposited on FTO compared to the yellow-orange color of CNON(Nb)-PC as can be seen in the photograph of Figure 3c. However, from the absorbance measurement, a clear conclusive result about the nitrogen content of the samples couldn't be obtained. Therefore, thermogravimetric (TG) analysis was used to compare the N content of the three different oxynitrides through the change of the mass with respect to temperature.

From the TG measurements shown in Figure 4, it can be observed that the three oxynitrides were stable up to a temperature of at least 200 °C. At higher temperatures the mass increased sharply due to the uptake of oxygen before decreasing again with the release of nitrogen. This qualitative behavior is in agreement with previous measurements reported in literature.[33] The onset temperature of the oxygen uptake was lower for the Nb enriched oxynitride with respect to the other two samples. The weight gain Δm after the complete oxidation of the oxynitride can be used to compare the N content of the three oxynitrides. From Figure 4, it can be clearly seen that the mass gain is more than three times higher for CNON(Nb)-PC compared to CNON-SS and CNON-PC. The mass difference is inversely proportional to the molar masses of the resulting oxides. However, since the difference in the molar masses of $CaNb_2O_6$ and $Ca_2Nb_2O_7$ obtained after TG is not significant, the higher mass difference of CNON(Nb)-PC is attributed to a higher N content than in CNON-SS and CNON-PC. Similar results were observed for $La(Ti_{1-x}Nb_x)O_2N$ where increasing the Nb content lowered the temperature at which oxygen uptake started and increased the N content in the oxynitrides.[34]

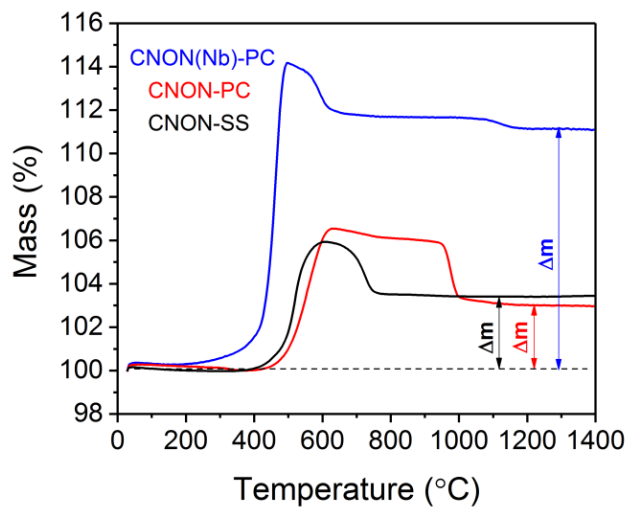

Figure 4. Thermogravimetric (TG) curves of the three calcium niobium oxynitride powders.

**Effect of Morphology on the PEC Performance of CNON.** CNON photoanodes were prepared by electrophoretic deposition of the oxynitride powders. The deposition was followed by a post necking treatment similar to that reported in literature for other oxynitrides in order to interconnect the particles and enhance the charge transport.[30] In brief, $TaCl_5$ was dropped on the as-prepared photoanode, then by annealing in air $Ta_2O_5$ connections were obtained. Post annealing in ammonia lead to the formation of Ta(O,N) that interconnected the CNON particles.

Figure S3 presents the current-voltage curves of CNON(Nb)-PC photoanodes in 0.5 M NaOH solution under chopped illumination with Xe lamp. The as-prepared photoanode showed a negligible photocurrent due to bad electrical contact between the particles. After necking with $Ta_2O_5$, the photocurrent increased to about 2 µA.cm$^{-2}$ at 1.23 V vs RHE. A significant enhancement in the photocurrent was achieved by annealing in $NH_3$ reaching 11.3 µA.cm$^{-2}$ at 1.23 V vs RHE. The post-necking treatment was also applied to CNON-SS and CNON-PC photoanodes.

Figure 5 shows the photoelectrochemical behavior of the oxynitrides with different morphologies. CNON(Nb)-PC showed the highest photocurrents in comparison to CNON-PC and CNON-SS. While photocurrents of about 10 and 14 µA.cm$^{-2}$ were achieved for CNON(Nb)-PC at 1.23 and 1.5 V vs RHE, respectively; CaNbO$_2$N-SS resulted in much lower photocurrents (4 and 6 µA.cm$^{-2}$ at 1.23 and 1.5 V vs RHE, respectively). CNON-PC resulted in the lowest photocurrents among the three different CNON photoanodes (≈1 and 3 µA.cm$^{-2}$ at 1.23 and 1.5 V vs RHE, correspondingly).

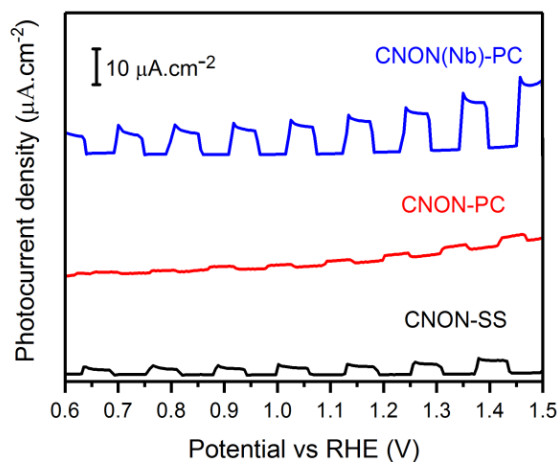

Figure 5. The potentiodynamic measurements of CNON-PC, CNON-SS, CNON(Nb)-PC acquired in NaOH (0.5M) solution under chopped light illumination with a Xe lamp.

The difference in morphology between the oxynitride powders, which was observed in the SEM images (Figure 2), is considered for a better understanding of the obtained photoactivities. The crystallinity of the photocatalysts is one of the major factors affecting the activity. Indeed, higher crystallinity leads to a better separation of the photogenerated charge carriers, thus leading to higher performance. However, CNON(Nb)-PC, which showed the smallest average grain size (lower crystallinity) among the three samples (see Figure 2f and Figure S1) exhibited higher photocurrents compared to the non-Nb enriched photoanodes with larger grain size. Hence, the photoactivity of CNON seems to be ruled mainly by factors other than crystallinity. In fact, it was reported that for LaTiO$_2$N photoanodes with different morphologies, the surface area had the most important effect on the photoactivity.[30] Thus, the surface area of the oxynitrides fabricated for this study was determined using BET. As shown in Figure 6, a clear trend was observed between the photocurrent densities of the CNON photoanodes and their BET surface areas. CNON(Nb)-PC had a surface area more than two and six times higher than CNON-SS and CNON-PC, respectively. The larger surface would result in more sites for hole injection into the electrolyte, that is more active sites for oxygen evolution.

Another advantage of the agglomerate structure of CNON(Nb)-PC is the low angle interparticle boundaries which allow the formation of densely packed films that would result in improved electron transfer. For CNON-SS particles, a closely packed film couldn't be formed due to the high angle interparticle boundaries as seen in the SEM image (Figure 2). One of the main factors to be considered for a rational design of particle-based photoelectrodes is that the particles should assemble into dense films providing large interparticle interface area.[30] This, in addition to the higher surface area, explain the improved performance of CNON(Nb)-PC electrodes compared to the others. In comparison, the low photoactivity of CNON-PC is attributed to the low surface area, and to the large nonporous cuboid particles shape which lead to poor charge transport as a result of its long migration distance.

Beside the morphological features, an important factor that influences the photoactivity of the oxynitride photoanodes is the visible light absorption properties and the N content. As discussed above, the high absorption background of both CNON-SS and CNON-PC shown in Figure 3b, suggests a greater amount of defects for these two oxynitrides compared to CNON(Nb)-PC. The defects act as recombination centers for the photogenerated e$^-$-h$^+$ pairs, thus reducing the photoactivity of the photoanodes. Additionally, the higher N content of CNON(Nb)-PC measured by TG improves the visible light absorption and thereby increases the PEC performance in comparison to the other two oxynitrides.

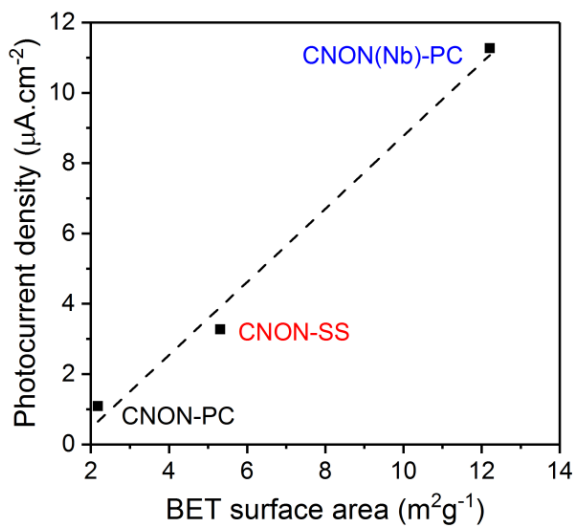

Figure 6. The photocurrent density of CNON oxynitrides measured at 1.23 V vs RHE (black squares) as a function of the BET surface area and a line shows the linear fitting of the data with slope of 1.04 µA.cm$^{-2}$ and R$^2$=0.989.

Indeed, it is expected that better performance could be achieved by optimizing the Nb enrichment in the oxide precursor. The XRD analysis of CNON(Nb)-PC did not show evidence of formation of secondary phases, whose presence may be expected in case that part of the excess Nb is not incorporated in the oxynitride. However, XRD analysis would not detect easily the presence of secondary phases with low crystallinity and/or very small average grain size. It was reported that up to 0.4 molar excess Nb improves PEC performance of BaNbO$_2$N, but for larger quantities of Nb excess the PEC performance decreases and the particle morphology changed showing cracks and the typical features of local segregation and agglomeration which were ascribed to secondary phases formed by the unreacted excess Nb. It was also shown that interestingly the 0.4 molar Nb excess corresponded to the value that resulted in a stoichiometric 1:1 ratio of the Ba and Nb content at the surface of the photoanode.[17] Future studies are planned to identify the optimal level of Nb excess in the synthesis of CNON.

**Effect of Co-catalyst and ALD layers on PEC Performance.** The slow kinetics of the oxygen evolution and the surface recombination of the photogenerated carriers usually limit the performance of photoanodes used for solar water splitting.[2, 14-15, 27] To tackle these issues, a co-catalyst is loaded on the semiconductor photoelectrode. The co-catalyst first captures the photogenerated holes thereby reducing the recombination of the photoinduced charge carriers. Then, the oxygen evolution reaction proceeds at the co-catalyst surface.

Among the different co-catalysts used to improve the photoactivity of photoanodes, Co-Pi is advantageous as it can be regenerated during the PEC experiments and also because of the ease of incorporating it into complex morphologies.[35] Several photoanodes, such as WO$_3$,[36] ZnO,[37] α-Fe$_2$O$_3$,[38-39] and W:BiVO$_4$[35], showed improved photoactivity when loaded with the Co-Pi co-catalyst. For the present study, Co-Pi was loaded on CNON photoanodes via photodeposition. This method was selected because it allows the deposition of the co-catalyst specifically in the sites where the photogenerated holes are more readily accessible,[37] thus, directly in the most active sites leading to better performance with reduced amount of Co-Pi.[40] The effect of Co-Pi loading was investigated using the best performing photoanode CNON(Nb)-PC. The photodeposition of Co-Pi was confirmed by SEM images. As clearly seen in Figure S1b, the co-catalyst nanoparticles (with average grain size of 10-15 nm) appeared homogeneously distributed on the CNON(Nb)-PC photoanode. The effect of Co-Pi on the photoactivity of the CaNbO$_2$N is assessed by comparing the PEC activity of bare CNON(Nb)-PC and CNON(Nb)-PC/Co-Pi photoanodes (Figure 7).

The photocurrent was enhanced with the photodeposition of Co-Pi to reach 19 and 25µA.cm$^{-2}$ at 1.23 and 1.5V vs RHE, respectively. This corresponds to an increment of about 90 and 80% of the photoactivity.

To complete the investigation of the potentials of CNON toward solar water splitting, we also probed the effect of two different passivation layers on the PEC activity. Several studies have shown that overlayers of oxides such as TiO$_2$,[41-42] Al$_2$O$_3$[43-46] and Ga$_2$O$_3$[47-48] have beneficial effects on the performance of photoanodes towards solar water splitting. These overlayers passivate the surface of the photoanodes by reducing the average density of surface defects. The surface defects, typically oxygen vacancies, act as recombination sites of the photogenerated charge carriers thus lowering the PEC performance of the photoelectrodes. For this study, overlayers of TiO$_2$ and Al$_2$O$_3$ were deposited by atomic layer deposition (ALD) on CNON(Nb)-PC/Co-Pi photoanodes.

As can be seen in Figure 7, the PEC performance of CNON(Nb)-PC/Co-Pi photoanodes remained almost the same after the deposition of TiO$_2$ overlayers with photocurrents of 17 and 24 µA.cm$^{-2}$ at 1.23 and 1.5 V vs RHE, respectively. Instead, after the deposition of Al$_2$O$_3$ an enhancement in the photocurrent up to 70 and 96 µA.cm$^{-2}$ at 1.23 and 1.5 V vs RHE was observed. This represents a 7-fold increase in the photocurrent compared to the bare photoanode. A similar observation was reported for hematite photoanodes whereby an increase by a factor of three of the photocurrent was achieved when passivating the surface with an alumina overlayer while no beneficial effect was observed with a TiO$_2$ layer. The improvement of the photocurrent was shown to be due to the passivation of the surface by the Al$_2$O$_3$ layer and not due to a catalytic effect. [46]

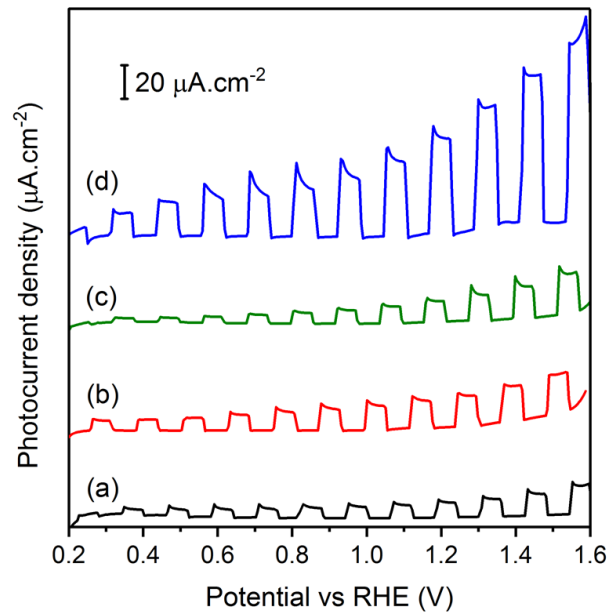

Figure 7. The potentiodynamic measurements of (a) bare CNON(Nb)-PC, (b) CNON(Nb)-PC/Co-Pi, (c) CNON(Nb)-PC/Co-Pi/TiO$_2$ and (d) CNON(Nb)-PC/Co-Pi/Al$_2$O$_3$ photoanodes.

For CNON(Nb)-PC, a significant surface density of defects, such as for instance oxygen vacancies, acting as recombination sites, may arise from the presence of reduced Nb ions near the surface. The deposition of an alumina overlayer may compensate these vacancies at the surface of the photoanodes. In addition, the alumina layer with more O$^{2-}$ ions compared to the oxynitride acts as an electron-rich layer which helps repelling of the photogenerated electrons from the surface thus reducing the extent of the surface recombination with holes.[49]

The photoelectrochemical stability of CNON photoanodes in NaOH (0.5 M) was studied by measuring the photocurrent at an applied potential of 1.23 V vs RHE for 30 min under illumination. Figure 8a shows the results of the stability tests performed for the bare CNON(Nb)-PC, CNON(Nb)-PC/Co-Pi, CNON(Nb)-PC/Co-Pi/TiO$_2$ and CNON(Nb)-PC/Co-Pi/Al$_2$O$_3$ photoanodes.

The rapid reduction of the photoactivity of the bare CNON(Nb)-PC photoanode is attributed to photooxidation of CNON, whereby the photogenerated holes promote the oxidation of N$^{3-}$ to N$_2$. This is a common problem reported for different oxynitrides.[10]

After the photodeposition of Co-Pi, a decrease of the photocurrent is still observed; that is mainly due to the incomplete coverage of the CNON surface by Co-Pi. However, after the modification of CNON surface with the TiO$_2$ layer, the photostability improved. The comparison of Figure 7 and 8a show that, though the TiO$_2$ layer did not improve the photocurrent nor reduce the recombination losses, this layer can indeed be used as a protective layer for the oxynitride. The TiO$_2$ coating allowed in fact to achieve and keep a stable value of photocurrent higher than that achieved using only the Co-Pi co-catalyst.

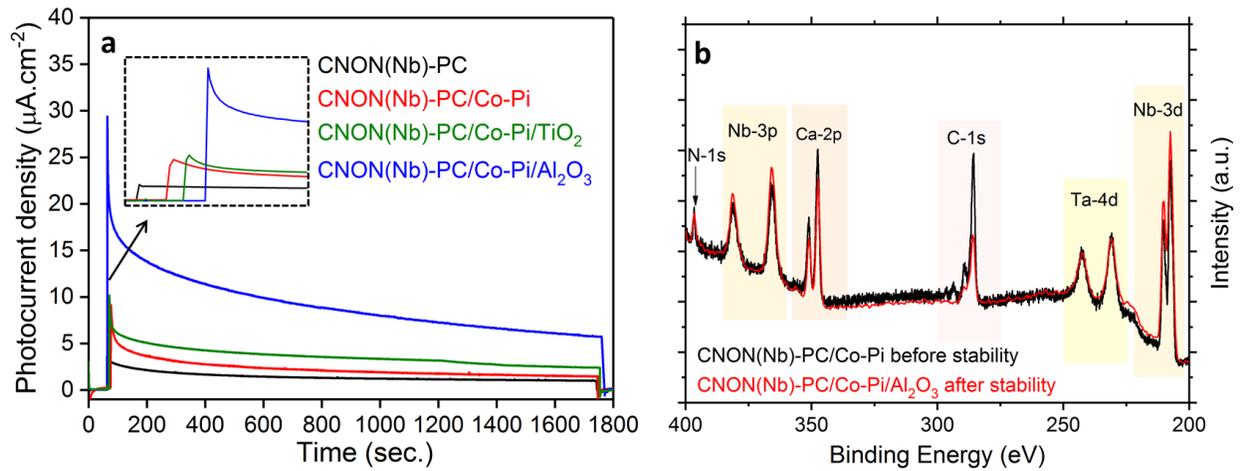

Figure 8. (a) Potentiostatic measurement showing the stability of bare CNON(Nb)-PC, CNON(Nb)-PC/Co-Pi, and CNON(Nb)-PC/Co-Pi/Al$_2$O$_3$ photoanodes for 30 min at 1.23 V vs RHE. The inset shows an enlarged view for the first 100sec of the stability measurement. (b) XPS spectra of CNON(Nb)-PC/Co-Pi before the stability test and CNON(Nb)-PC/Co-Pi/Al$_2$O$_3$ after the stability test.

Differently, the CNON(Nb)-PC/Co-Pi/Al$_2$O$_3$ photoanode showed remarkably better performance compared to the uncoated sample. The initial photocurrent density was 3 to 4 times higher than that measured for the Co-Pi/TiO$_2$ coated samples. The photocurrent however decreased to a half of the initial value within the first 10 minutes. This effect is mainly due to the dissolution of Al$_2$O$_3$ in the NaOH electrolyte, as confirmed by the XPS analyses shown in Figure S4.

This observation is in agreement with what one would expect considering the etching properties of alumina reported in literature.[50] In an alkaline solution with pH of 13, a 2 nm thick alumina layer is expected to be dissolved at room temperature in about 3 minutes. Nevertheless, an alumina overlayer was reported to be beneficial also to boost the photoactivity of hematite photoanodes.[46] In that study 1 M NaOH was used as the electrolyte with similar value of pH as in the present investigation, however the alumina overlayers was reported to be sufficiently stable for at least 30 minutes.

For our sample XPS suggests that the beneficial effect of the Al$_2$O$_3$ overlayer is limited by its solubility in the alkaline medium used for this study. However, even though this overlayer dissolved during the PEC measurement, it can be observed in Figure 8a that CNON(Nb)-PC/Co-Pi/Al$_2$O$_3$ still maintained a photoactivity about 50% higher than the CNON(Nb)-PC/Co-Pi photoanode without a passivation layer.

To gain more insights into the observed enhancement of photocurrent density, XPS measurements were conducted to investigate the surface chemistry of the Co-Pi loaded CNON(Nb)-PC photoanodes with and without passivation layer (Figure 8b). XPS indicates that 1) the as prepared sample had a significant Nb surface enrichment (Ca/Nb = 0.77), 2) the Nb enrichment increased after the deposition of the Al$_2$O$_3$ overlayer (Ca/Nb = 0.48) but 3) the Nb content remained stable after PEC tests (Ca/Nb = 0.53).

The surface Nb enrichment of the as prepared sample is likely to arise from the thermal treatments required for the fabrication of the photoanode which results in the segregation of Nb to the surface. The further enrichment detected after Al$_2$O$_3$ deposition could be due to the additional thermal treatment the sample undergoes during ALD. However, the PEC characterization of samples that underwent the same thermal treatments but without the Al$_2$O$_3$ deposition showed no enhancement of photocurrent. We thus conclude that the Nb surface enrichment is not the reason of the improved performance of CNON(Nb)-PC/Co-Pi/Al$_2$O$_3$. Besides, ascribing the increased PEC performance to Nb segregation would not agree with previous measurements on BaNbO$_2$N[17] showing that a 1:1 Ca/Nb ratio at the surface leads to the highest photoactivity. Actually, as previously pointed out, assuming similar behavior for CNON and BaNbO$_2$N, the XPS measurements reported here suggest that the performance of CNON(Nb)-PC/Co-Pi/Al$_2$O$_3$ could be further improved by optimizing the chemical composition of the surface to reach a 1:1 Ca/Nb ratio.

One alternative possibility to explain the enhanced photoactivity of CNON(Nb)-PC/Co-Pi/Al$_2$O$_3$ is that the presence of the O-rich alumina layer at the surface favors the oxidation of the Nb ions decreasing the amount of Nb$^{4+}$ that act as recombination centers. However, also this scenario can be ruled out. Figure S5 in fact shows that the Nb$^{4+}$ and Nb$^{5+}$ components contribute equally to the Nb 3d XPS spectrum of the photoanodes with and without Al$_2$O$_3$ overlayer.

Finally, we suggest that the observed significant and prolonged improvement of the photocurrent density could be attributed to the increased roughness of the surface after the dissolution of the passivation layer. An increased surface roughness would result in an increased semiconductor/electrolyte interface that would in turn improve the PEC performance of the photoanodes. For instance, it was shown that the leaching of Cr from outermost surface of Fe$_2$O$_3$ modified by a Fe$_{20}$Cr$_{40}$Ni$_{40}$O$_x$ layer increases the surface roughness.[51] To verify this hypothesis, the electrochemical active surface area of the CNON(Nb)-PC/Co-Pi and CNON(Nb)-PC/Co-Pi/Al$_2$O$_3$ after PEC experiments was evaluated. From the change of current in the non-Faradaic region, which corresponds only to the charge and discharge of the electric double layer, with respect to the scan rate, the electrochemical area can be estimated. Cyclic voltammetry (CV) measurements were performed at different scan rates for the two photoanodes. Figure 9 shows the values of current measured from the CV scans versus the scan rate for the CNON(Nb)-PC/Co-Pi with and without Al$_2$O$_3$ overlayer after the stability test shown in Figure 8a.

The slope of the fitted lines corresponds to the total Helmholtz capacitance ($C_H$) of the photoanode:

$$C_H = C_{H,sp} \times A$$

where $C_{H,sp}$ is the specific Helmholtz capacitance (F.cm$^{-2}$) and $A$ is the electrochemical active surface area (in cm$^2$). There are no previous reports on the value of $C_{H,sp}$ for CNON, however, since this value is the same for both photoanodes, the difference in $C_H$ (slope of the fitted lines in Figure 9) can only be due to the change of the surface area. For CNON(Nb)-PC/Co-Pi/Al$_2$O$_3$ the slope of the curve in Figure 9 is 2.3 times higher than that calculated for CNON(Nb)-PC/Co-Pi. This means that after the stability test, the dissolution of Al$_2$O$_3$ results in an increased surface area of the photoanode.

By comparing the photocurrent values for both photoanodes after the potentiostatic measurements shown in Figure 8a, an enhancement in the photocurrent by a factor of 2.8 can be observed for the Al$_2$O$_3$-modified sample.

This enhancement is similar to the estimated increase of the electrochemical active surface area. Therefore, we conclude that even when the Al$_2$O$_3$ overlayer is dissolved, a prolonged photocurrent enhancement is observed mainly due to the increased surface area.

Further studies are required to investigate the use of other overlayer materials and to understand why Al$_2$O$_3$ and TiO$_2$ have different functionalities.

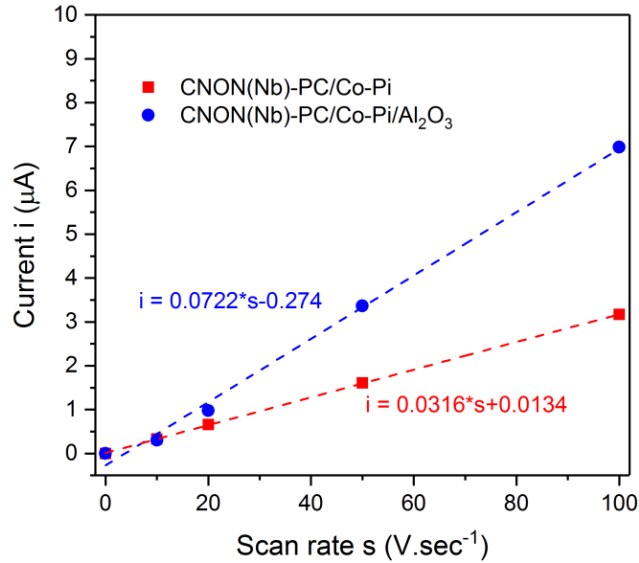

Figure 9. Plot of current as a function of scan rate for CNON(Nb)-PC/Co-Pi and CNON(Nb)-PC/Co-Pi/Al$_2$O$_3$ photoanodes after the photostability measurement.

## 4. CONCLUSION

In summary, we showed in this work the first photoelectrochemical water splitting measurements for CaNbO$_2$N (CNON) photoanodes. CNON synthesized from different oxide precursors resulted in different performances of the photoanodes, mainly due to the different morphologies and surface areas. The Nb-enriched oxide resulted in the best performing photoanode, which was further improved by the photodeposition of Co-Pi co-catalyst and the ALD deposition of Al$_2$O$_3$ overlayer. An improved PEC activity of the CNON(Nb)-PC/Co-Pi/Al$_2$O$_3$ photoanode was observed even after the Al$_2$O$_3$ layer was completely dissolved in the alkaline electrolyte. This is attributed to the increased surface roughness of the photoanodes after the dissolution of Al$_2$O$_3$ layer. The modification of CNON(Nb)-PC/Co-Pi photoanodes with an ALD layer of TiO$_2$ didn't result in an enhanced PEC activity; however it improved the stability of the photoanodes.

These results encourage further investigations using other passivation layers for an improved photoactivity and stability of the CNON photoanodes.

## ASSOCIATED CONTENT

### Supporting Information

Lattice parameters of calcium niobium oxides and oxynitride, SEM images of CNON(PC)-Nb photoanode before and after loading of Co-Pi co-catalyst, Tauc plots of CNON(PC)-Nb, CNON-PC and CNON-SS, potentiodynamic scans of before and after post treatment, XPS of Al-2p and Nb-3d peaks. This material is available free of charge via the Internet at http://pubs.acs.org.

## AUTHOR INFORMATION

### Corresponding Author

* Emails: Daniele.pergolesi@psi.ch, Thomas.lippert@psi.ch

### Present Addresses

§ (W.Si) Key Laboratory of Advanced Ceramics and Machining Technology, Ministry of Education, School of Materials Science and Engineering, Tianjin University, Tianjin 300072, P.R. China.

## ACKNOWLEDGMENT

This research was supported by the Paul Scherrer Institut and the NCCR MARVEL funded by the Swiss National Science Foundation. The Swiss Excellence Governmental Scholarship is gratefully acknowledged for the financial support of this work. Fatima Haydous was a Swiss Government Excellence Scholarship holder for the academic years 2014-2017 (ESKAS No. 2014.0282). The authors would like to thank Dr. Bilal Gökce from the University of Duisburg-Essen for the support with the BET measurements and Dr. David Tilley from the University of Zurich for the help with the atomic layer deposition.

## REFERENCES

1. Lewis, N. S. Toward Cost-Effective Solar Energy Use. *Science (N.Y.)* **2007**, *315*, 798-801.
2. Gujral, S. S.; Simonov, A. N.; Higashi, M.; Fang, X.-Y.; Abe, R.; Spiccia, L. Highly Dispersed Cobalt Oxide on Taon as Efficient Photoanodes for Long-Term Solar Water Splitting. *ACS Catal.* **2016**, *6*, 3404-3417.
3. Allam, N. K.; Shaheen, B. S.; Hafez, A. M. Layered Tantalum Oxynitride Nanorod Array Carpets for Efficient Photoelectrochemical Conversion of Solar Energy: Experimental and Dft Insights. *ACS Appl. Mater. Interfaces* **2014**, *6*, 4609-4615.
4. Kubota, J.; Domen, K. Photocatalytic Water Splitting Using Oxynitride and Nitride Semiconductor Powders for Production of Solar Hydrogen. *Electrochem. Soc. Interface* **2013**, *22*, 57-62.
5. Maeda, K.; Higashi, M.; Siritanaratkul, B.; Abe, R.; Domen, K. SrNbO$_2$N as a Water-Splitting Photoanode with a Wide Visible-Light Absorption Band. *J. Am. Chem. Soc.* **2011**, *133*, 12334-12337.
6. Pichler, M.; Szlachetko, J.; Castelli, I. E.; Marzari, N.; Döbeli, M.; Wokaun, A.; Pergolesi, D.; Lippert, T. Determination of Conduction and Valence Band Electronic Structure of Latioxny Thin Film. *ChemSusChem* **2017**, *10*, 2099-2106.
7. Maegli, A. E.; Pokrant, S.; Hisatomi, T.; Trottmann, M.; Domen, K.; Weidenkaff, A. Enhancement of Photocatalytic Water Oxidation by the Morphological Control of LaTiO$_2$N and Cobalt Oxide Catalysts. *J. Phys. Chem. C* **2014**, *118*, 16344-16351.
8. Singh, R. B.; Matsuzaki, H.; Suzuki, Y.; Seki, K.; Minegishi, T.; Hisatomi, T.; Domen, K.; Furube, A. Trapped State Sensitive Kinetics in LaTiO$_2$N Solid Photocatalyst with and without Cocatalyst Loading. *J. Am. Chem. Soc.* **2014**, *136*, 17324-17331.
9. Yamakata, A.; Kawaguchi, M.; Nishimura, N.; Minegishi, T.; Kubota, J.; Domen, K. Behavior and Energy States of Photogenerated Charge Carriers on Pt- or CoO$_x$-Loaded Latio2n Photocatalysts: Time-Resolved Visible to Mid-Infrared Absorption Study. *J. Phys. Chem. C* **2014**, *118*, 23897-23906.
10. Higashi, M.; Domen, K.; Abe, R. Fabrication of an Efficient BaTaO$_2$N Photoanode Harvesting a Wide Range of Visible Light for Water Splitting. *J. Am. Chem. Soc.* **2013**, *135*, 10238-10241.
11. Higashi, M.; Yamanaka, Y.; Tomita, O.; Abe, R. Fabrication of Cation-Doped BaTaO$_2$N Photoanodes for Efficient Photoelectrochemical Water Splitting under Visible Light Irradiation. *APL Mater.* **2015**, *3*, 104418.
12. Wang, C.; Hisatomi, T.; Minegishi, T.; Wang, Q.; Zhong, M.; Katayama, M.; Kubota, J.; Domen, K. Synthesis of Nanostructured BaTaO$_2$N Thin Films as Photoanodes for Solar Water Splitting. *J. Phys. Chem. C* **2016**, *120*, 15758–15764.
13. Abe, R.; Higashi, M.; Domen, K. Facile Fabrication of an Efficient Oxynitride TaON Photoanode for Overall Water Splitting into H$_2$ and O$_2$ under Visible Light Irradiation. *J. Am. Chem. Soc.* **2010**, *132*, 11828-11829.
14. Higashi, M.; Domen, K.; Abe, R. Fabrication of Efficient Taon and Ta$_3$N$_5$ Photoanodes for Water Splitting under Visible Light Irradiation. *Energy Environ. Sci.* **2011**, *4*, 4138-4147.
15. Higashi, M.; Domen, K.; Abe, R. Highly Stable Water Splitting on Oxynitride TaON Photoanode System under Visible Light Irradiation. *J. Am. Chem. Soc.* **2012**, *134*, 6968-6971.
16. Siritanaratkul, B.; Maeda, K.; Hisatomi, T.; Domen, K. Synthesis and Photocatalytic Activity of Perovskite Niobium Oxynitrides with Wide Visible-Light Absorption Bands. *ChemSusChem* **2011**, *4*, 74-78.
17. Seo, J.; Moriya, Y.; Kodera, M.; Hisatomi, T.; Minegishi, T.; Katayama, M.; Domen, K. Photoelectrochemical Water Splitting on Particulate ANbO$_2$N (A = Ba, Sr) Photoanodes Prepared from Perovskite-Type ANbO$_3$. *Chem. Mater.* **2016**, *28*, 6869-6876.
18. Castelli, I. E.; Landis, D. D.; Thygesen, K. S.; Dahl, S.; Chorkendorff, I.; Jaramillo, T. F.; Jacobsen, K. W. New Cubic Perovskites for One- and Two-Photon Water Splitting Using the Computational Materials Repository. *Energy Environ. Sci.* **2012**, *5*, 9034-9043.
19. Maeda, K.; Higashi, M.; Siritanaratkul, B.; Abe, R.; Domen, K. SrNbO$_2$N as a Water-Splitting Photoanode with a Wide Visible-Light Absorption Band. *J. Am. Chem. Soc.* **2011**, *133*, 12334-12337.
20. Hisatomi, T.; Katayama, C.; Moriya, Y.; Minegishi, T.; Katayama, M.; Nishiyama, H.; Yamada, T.; Domen, K. Photocatalytic Oxygen Evolution Using BaNbO$_2$N Modified with Cobalt Oxide under Photoexcitation up to 740 nm. *Energy Environ. Sci.* **2013**, *6*, 3595-3599.

21. Landsmann, S.; Surace, Y.; Trottmann, M.; Dilger, S.; Weidenkaff, A.; Pokrant, S. Controlled Design of Functional Nano-Coatings: Reduction of Loss Mechanisms in Photoelectrochemical Water Splitting. *ACS Appl. Mater. Interfaces* **2016**, *8,* 12149–12157.
22. Dabirian, A.; Van de Krol, R. High-Temperature Ammonolysis of Thin Film $Ta_2O_5$ Photoanodes: Evolution of Structural, Optical, and Photoelectrochemical Properties. *Chem. Mater.* **2015**, *27,* 708-715.
23. Si, W.; Pergolesi, D.; Haydous, F.; Fluri, A.; Wokaun, A.; Lippert, T. Investigating the Behavior of Various Cocatalysts on $LaTaON_2$ Photoanode for Visible Light Water Splitting. *Phys. Chem. Chem. Phys.* **2017**, *19,* 656-662.
24. Kato, H.; Ueda, K.; Kobayashi, M.; Kakihana, M. Photocatalytic Water Oxidation under Visible Light by Valence Band Controlled Oxynitride Solid Solutions $LaTaON_2$-$SrTiO_3$. *J. Mater. Chem. A* **2015**, *3,* 11824-11829.
25. Oehler, F.; Naumann, R.; Köferstein, R.; Hesse, D.; Ebbinghaus, S. G. Photocatalytic Activity of $CaTaO_2N$ Nanocrystals Obtained from a Hydrothermally Synthesized Oxide Precursor. *Mater. Res. Bull.* **2016**, *73,* 276-283.
26. Lim, Y.-F.; Chua, C. S.; Lee, C. J. J.; Chi, D. Sol-Gel Deposited $Cu_2O$ and Cuo Thin Films for Photocatalytic Water Splitting. *Phys. Chem. Chem. Phys.* **2014**, *16,* 25928-25934.
27. Hisatomi, T.; Kubota, J.; Domen, K. Recent Advances in Semiconductors for Photocatalytic and Photoelectrochemical Water Splitting. *Chem. Soc. Rev.* **2014**, *43,* 7520-7535.
28. Arunachalam, P.; Al-Mayouf, A.; Ghanem, M. A.; Shaddad, M. N.; Weller, M. T. Photoelectrochemical Oxidation of Water Using $La(Ta, Nb)O_2N$ Modified Electrodes. *Int. J. Hydrogen Energy* **2016**, *41,* 11644-11652.
29. Jang, J.-W.; Chun, D.; Ye, Y.; Lin, Y.; Yao, X.; Thorne, J.; Liu, E.; McMahon, G.; Zhu, J.; Javey, A.; et al. Enabling Unassisted Solar Water Splitting by Iron Oxide and Silicon. *Nat. Commun.* **2015**, *6,* 7447.
30. Landsmann, S.; Maegli, A. E.; Trottmann, M.; Battaglia, C.; Weidenkaff, A.; Pokrant, S. Design Guidelines for High-Performance Particle-Based Photoanodes for Water Splitting: Lanthanum Titanium Oxynitride as a Model. *ChemSusChem* **2015**, *8,* 3451 –3458.
31. Zhang, F.; Yamakata, A.; Maeda, K.; Moriya, Y.; Takata, T.; Kubota, J.; Teshima, K.; Oishi, S.; Domen, K. Cobalt-Modified Porous Single-Crystalline $LaTiO_2N$ for Highly Efficient Water Oxidation under Visible Light. *J. Am. Chem. Soc.* **2012**, *134,* 8348-8351.
32. Feng, J.; Luo, W.; Fang, T.; Lv, H.; Wang, Z.; Gao, J.; Liu, W.; Yu, T.; Li, Z.; Zou, Z. Highly Photo-Responsive $LaTiO_2N$ Photoanodes by Improvement of Charge Carrier Transport among Film Particles. *Adv. Funct. Mater.* **2014**, *24,* 3535-3542.
33. Le Gendre, L.; Marchand, R.; Laurent, Y. A New Class of Inorganic Compounds Containing Dinitrogen-Metal Bonds. *J. Eur. Ceram. Soc.* **1997**, *17,* 1813-1818.
34. Yoon, S.; Maegli, A. E.; Eyssler, A.; Trottmann, M.; Hisatomi, T.; Leroy, C. M.; Grätzel, M.; Weidenkaff, A. Synthesis and Characterization of $La(Ti,Nb)(O,N)_3$ for Photocatalytic Water Oxidation. *Energy Procedia* **2012**, *22,* 41-47.
35. Zhong, D. K.; Choi, S.; Gamelin, D. R. Near-Complete Suppression of Surface Recombination in Solar Photoelectrolysis by "Co-Pi" Catalyst-Modified $W:BiVO_4$. *J. Am. Chem. Soc.* **2011**, *133,* 18370-18377.
36. Seabold, J. A.; Choi, K.-S. Effect of a Cobalt-Based Oxygen Evolution Catalyst on the Stability and the Selectivity of Photo-Oxidation Reactions of a $WO_3$ Photoanode. *Chem. Mater.* **2011**, *23,* 1105-1112.
37. Steinmiller, E. M. P.; Choi, K.-S. Photochemical Deposition of Cobalt-Based Oxygen Evolving Catalyst on a Semiconductor Photoanode for Solar Oxygen Production. *Proc. Natl. Acad. Sci.* **2009**, *106,* 20633-20636.
38. Zhong, D. K.; Sun, J.; Inumaru, H.; Gamelin, D. R. Solar Water Oxidation by Composite Catalyst/A-$Fe_2O_3$ Photoanodes. *J. Am. Chem. Soc.* **2009**, *131,* 6086-6087.
39. Zhong, D. K.; Cornuz, M.; Sivula, K.; Grätzel, M.; Gamelin, D. R. Photo-Assisted Electrodeposition of Cobalt–Phosphate (Co–Pi) Catalyst on Hematite Photoanodes for Solar Water Oxidation. *Energy Environ. Sci.* **2011**, *4,* 1759-1764.
40. McDonald, K. J.; Choi, K.-S. Photodeposition of Co-Based Oxygen Evolution Catalysts on a-$Fe_2O_3$ Photoanodes. *Chem. Mater.* **2011**, *23,* 1686-1693.
41. Yang, X.; Liu, R.; Du, C.; Dai, P.; Zheng, Z.; Wang, D. Improving Hematite-Based Photoelectrochemical Water Splitting with Ultrathin $TiO_2$ by Atomic Layer Deposition. *ACS Appl. Mater. Interfaces* **2014**, *6,* 12005-12011.
42. Li, C.; Hisatomi, T.; Watanabe, O.; Nakabayashi, M.; Shibata, N.; Domen, K.; Delaunay, J.-J. Positive Onset Potential and Stability of $Cu_2O$-Based Photocathodes in Water Splitting by Atomic Layer Deposition of a $Ga_2O_3$ Buffer Layer. *Energy Environ. Sci.* **2015**, *8,* 1493-1500.
43. Zeng, M.; Peng, X.; Liao, J.; Wang, G.; Li, Y.; Li, J.; Qin, Y.; Wilson, J.; Song, A.; Lin, S. Enhanced Photoelectrochemical Performance of Quantum Dot-Sensitized $TiO_2$ Nanotube Arrays with $Al_2O_3$ Overcoating by Atomic Layer Deposition. *Phys. Chem. Chem. Phys.* **2016**, *18,* 17404-17413.
44. Neufeld, O.; Yatom, N.; Caspary Toroker, M. A First-Principles Study on the Role of an $Al_2O_3$ Overlayer on $Fe_2O_3$ for Water Splitting. *ACS Catal.* **2015**, *5,* 7237-7243.
45. Roelofs, K. E.; Brennan, T. P.; Dominguez, J. C.; Bailie, C. D.; Margulis, G. Y.; Hoke, E. T.; McGehee, M. D.; Bent, S. F. Effect of $Al_2O_3$ Recombination Barrier Layers Deposited by Atomic Layer Deposition in Solid-State CdS Quantum Dot-Sensitized Solar Cells. *J. Phys. Chem. C* **2013**, *117,* 5584-5592.
46. Le Formal, F.; Tetreault, N.; Cornuz, M.; Moehl, T.; Gratzel, M.; Sivula, K. Passivating Surface States on Water Splitting Hematite Photoanodes with Alumina Overlayers. *Chem. Sci.* **2011**, *2,* 737-743.
47. Barroso, M.; Mesa, C. A.; Pendlebury, S. R.; Cowan, A. J.; Hisatomi, T.; Sivula, K.; Grätzel, M.; Klug, D. R.; Durrant, J. R. Dynamics of Photogenerated Holes in Surface Modified A-$Fe_2O_3$ Photoanodes for Solar Water Splitting. *Proc. Natl. Acad. Sci.* **2012**, *109,* 15640-15645.
48. Steier, L.; Herraiz-Cardona, I.; Gimenez, S.; Fabregat-Santiago, F.; Bisquert, J.; Tilley, S. D.; Grätzel, M. Understanding the Role of Underlayers and Overlayers in Thin Film Hematite Photoanodes. *Adv. Funct. Mater.* **2014**, *24,* 7681-7688.
49. Le Formal, F.; Sivula, K.; Grätzel, M. The Transient Photocurrent and Photovoltage Behavior of a Hematite Photoanode under Working Conditions and the Influence of Surface Treatments. *J. Phys. Chem. C* **2012**, *116,* 26707-26720.
50. Sun, K. G.; Li, Y. V.; Saint John, D. B.; Jackson, T. N. Ph-Controlled Selective Etching of $Al_2O_3$ over Zno. *ACS Appl. Mater. Interfaces* **2014**, *6,* 7028-7031.
51. Bärtsch, M.; Sarnowska, M.; Krysiak, O.; Willa, C.; Huber, C.; Pillatsch, L.; Reinhard, S.; Niederberger, M. Multicomposite Nanostructured Hematite–Titania Photoanodes with Improved Oxygen Evolution: The Role of the Oxygen Evolution Catalyst. *ACS Omega* **2017**, *2,* 4531-4539.

TOC Graphic

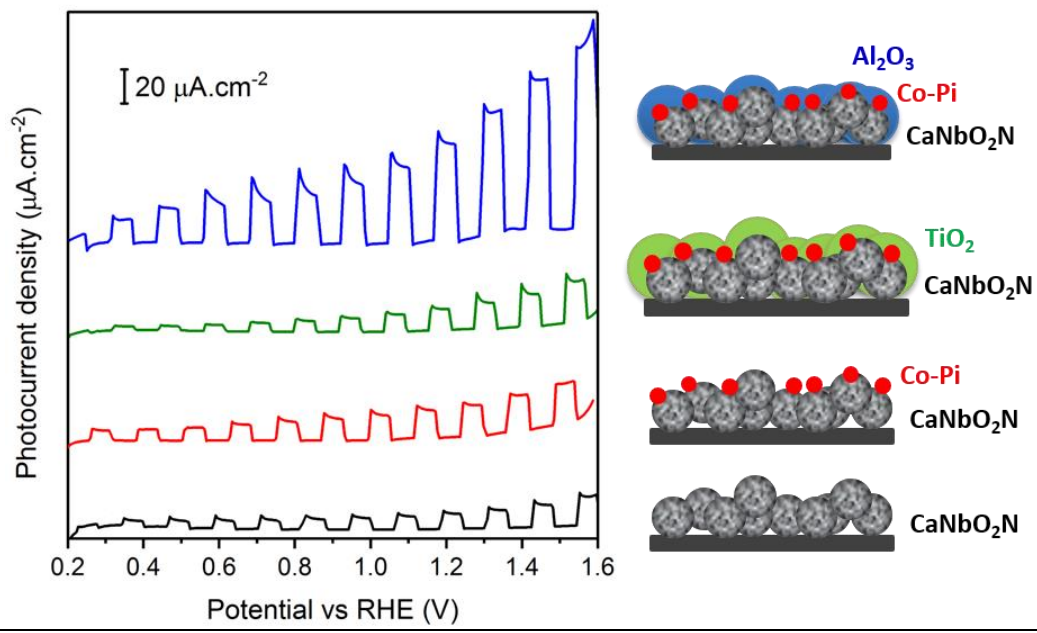

# Supporting Information

Table S1. Lattice Parameters of Ca$_2$Nb$_2$O$_7$ (ICSD Coll.Code: 26010), CaNb$_2$O$_6$ (ICSD Coll.Code:15208), and CaNbO$_2$N (ICSD Coll.Code: 55396).

| Material | Crystal Structure | Space Group | Lattice Parameters Å | | |
|---|---|---|---|---|---|
| | | | a | b | c |
| Ca$_2$Nb$_2$O$_7$ | monoclinic | P1121 | 7.697 | 13.385 | 5.502 |
| CaNb$_2$O$_6$ | orthorhombic | Pbcn | 14.926 | 5.752 | 5.204 |
| CaNbO$_2$N | orthorhombic | Pnma | 5.64051 | 7.90711 | 5.55508 |

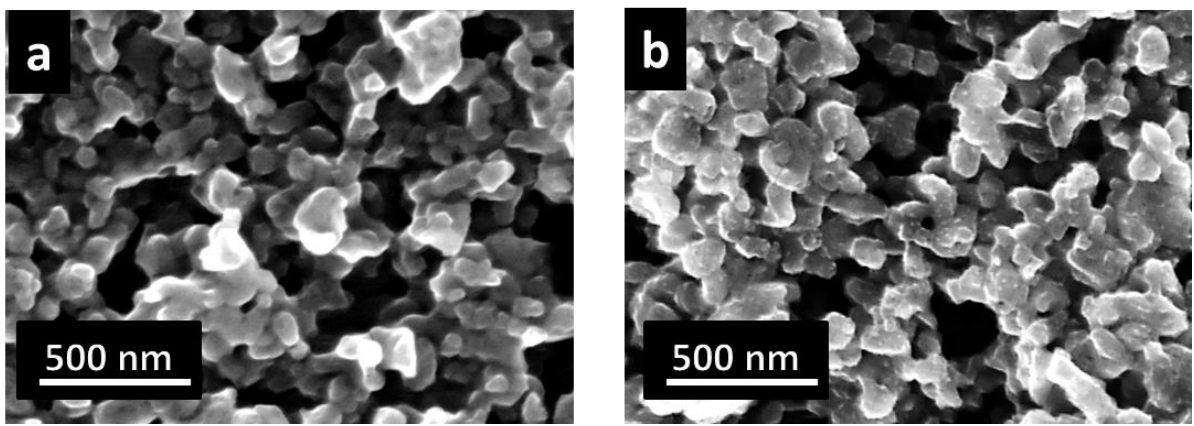

Figure S5. SEM images of CNON(Nb)-PC/Co-Pi photoanode (a) before and (b) after Co-Pi photodeposition.

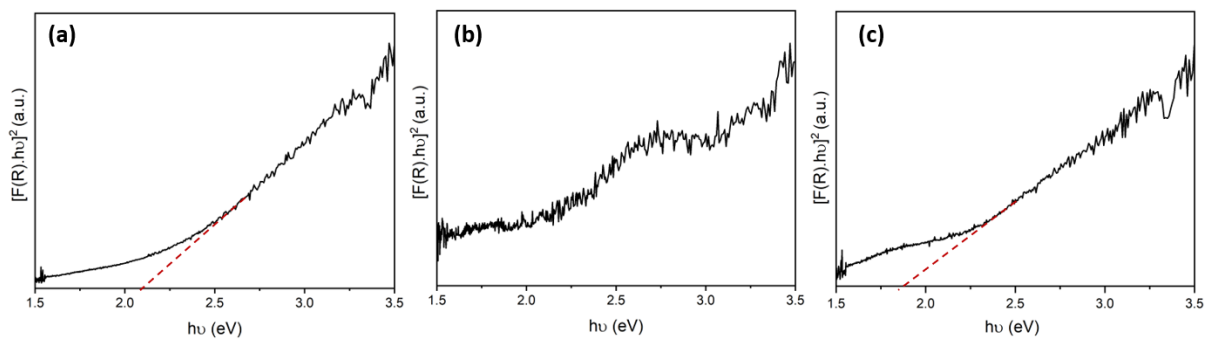

Figure S6. Tauc plots of (a) CNON(Nb)-PC, (b) CNON-PC and (c) CNON-SS.

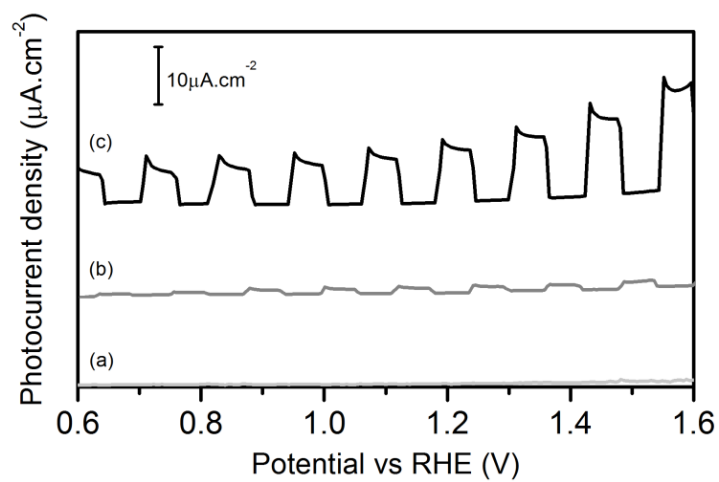

Figure S7. Potentiodynamic scans of CNON(Nb)-PC photoanodes (a) before post-treatment, (b) after post necking with TaCl5 and annealing in air, and (c) after annealing in NH3 following the post necking procedure in 0.5 M NaOH solution under chopped illumination with Xe lamp.

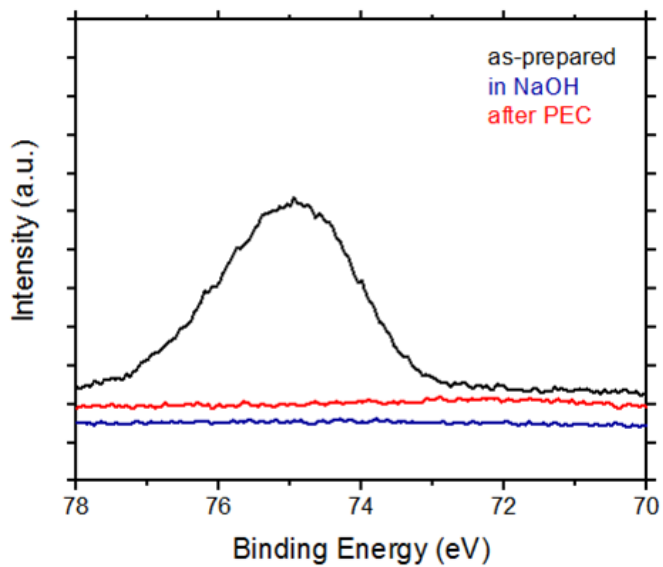

Figure S8. XPS spectra of Al-2p for CNON(Nb)-PC/Co-Pi/Al$_2$O$_3$ photoanode as-prepared, soaked in NaOH for 30 min, and after the photostability test.

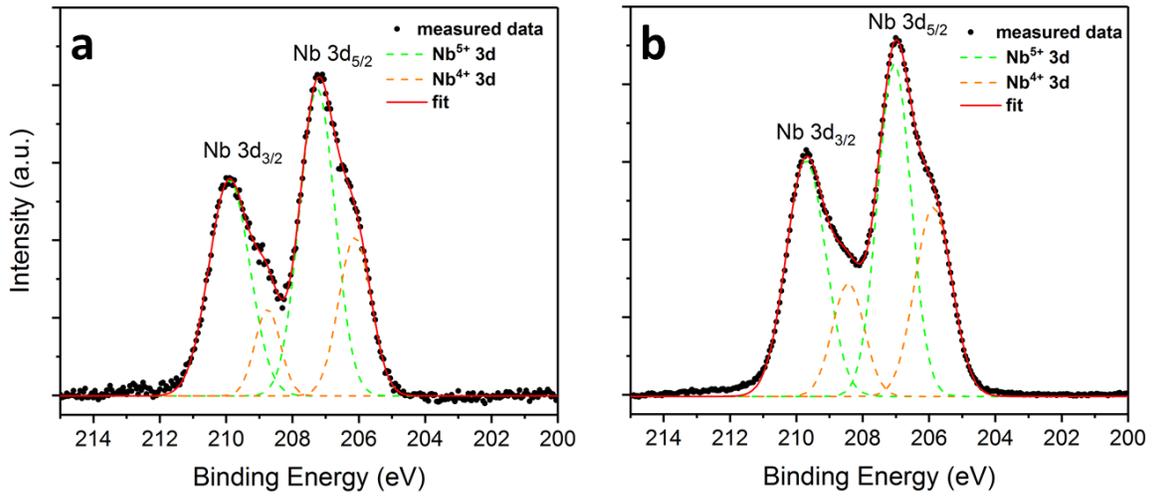

Figure S9. XPS spectra for Nb 3d peaks for (a) as-prepared CNON(Nb)-PC/Co-Pi and (b) CNON(Nb)-PC/Co-Pi/$Al_2O_3$ after PEC photoanodes.

17